\documentclass{moriond}

\newcommand{\Photo}{}

\begin{document}

\begin{flushright}
MS-TP-21-17\\
LTH 1261\\
\end{flushright}

\vspace*{4cm}
\title{Slepton and Electroweakino pair production with aNNLO+NNLL precision}

\author{Juri Fiaschi$^a$ and Michael Klasen$^b$}

\address{$^a$ Department of Mathematical Sciences, University of Liverpool, Liverpool L69 3BX, United Kingdom}
\address{$^b$ Institut für Theoretische Physik, Westfälische Wilhelms-Universität Münster, Wilhelm-Klemm-Straße 9, 48149 Münster, Germany}

\maketitle\abstracts{
The level of experimental precision that will be achieved with LHC Run-III data and in the forthcoming High Luminosity stage calls for equally accurate theoretical predictions to compare with. Here we present updated cross section calculations for the electroweak production of SUSY particles at the LHC with aNNLO+NNLL (approximate next-to-next-to-leading-order plus next-to-next-to-leading-logarithmic) accuracy. Results are shown for slepton pair production and for electroweakinos pair production in their mostly higgsino or gaugino configuration, finding a further significant reduction of the factorisation and renormalisation scale dependence that stabilises the predictions to the permil level.
}

\section{Introduction}

The search for suspersymmetry (SUSY) is one of the most fascinating ongoing investigations at the LHC.
The Minimal Supersymmetric Standard Model (MSSM) provides an elegant framework for a possible realization of a supersymmetric extension of the Standard Model (SM), solving few of the drawbacks of the latter, as it provides a natural explanation for the protection of the Higgs boson mass from radiative corrections, it predicts the unification of strong and electroweak forces at high scales, and it also contains Dark Matter candidates which can naturally explain modern cosmological observations.

The large amount of data that is being collected at the LHC is converted into precise measurements of a wide range of possible SUSY signatures by the analysis of the CMS and ATLAS experimental groups.
These measurements are in turn translated into increasingly stringent limits on the masses of the SUSY particles, under the condition that the accuracy of the theoretical predictions match the experimental precision.

Radiative corrections at next-to-leading order (NLO) and beyond have been calculated for many SUSY processes, and ad hoc techniques have been developed for the treatment of the logarithmic terms that appear in the cross section perturbative expansion and that can be large in specific kinematical regions.
The effects of the resummation of these terms at next-to-leading logarithmic (NLL) accuracy and beyond have been investigated thoroughly in the past for several SUSY processes.

In this work, we present updated predictions for the cross sections of sleptons pair~\cite{Fiaschi:2019zgh} and electroweakinos production~\cite{Fiaschi:2020udf} including threshold NNLL corrections matched at aNNLO, which contains SUSY corrections only at NLO, since they are not known beyond this order.

\section{Slepton pair production at aNNLO+NNLL}

We present here the results for the cross section of pair produced left-handed first or second generation sleptons.
Predictions for the case of third generation sleptons or for different chiralities of the final state particles can be obtained by a simple consistent rescaling of the results.

The most recent exclusion limits by ATLAS and CMS derived from the 139 fb$^{-1}$ integrated luminosity data set, force the mass of these particles above 700 GeV~\cite{Aad:2019vnb,Sirunyan:2020eab}.

\begin{figure}
\centering
\includegraphics[width=0.43\textwidth]{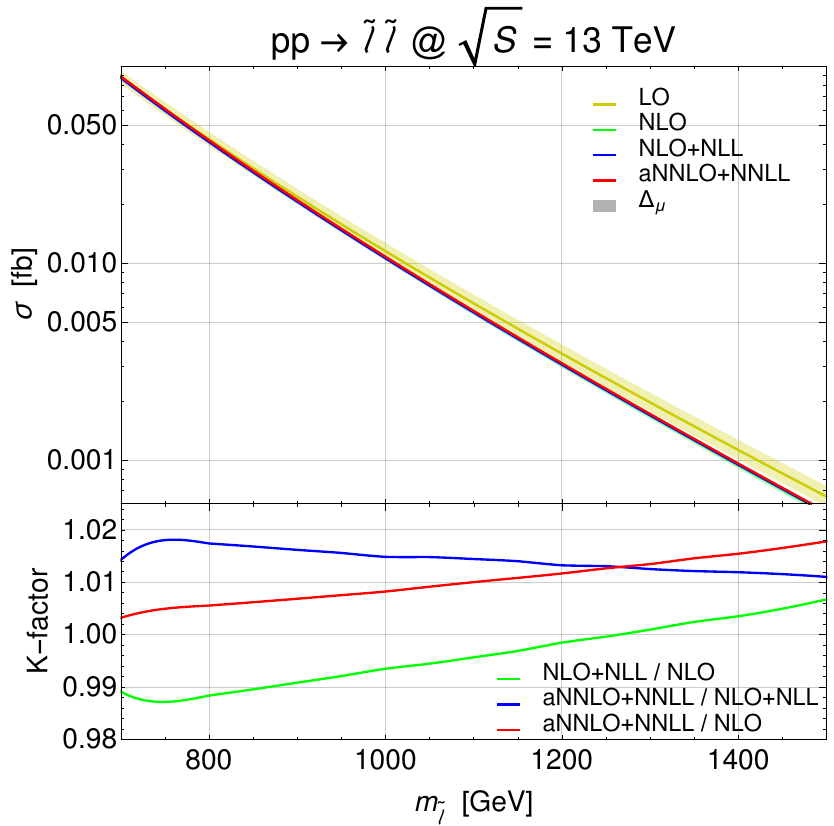}
\includegraphics[width=0.56\textwidth]{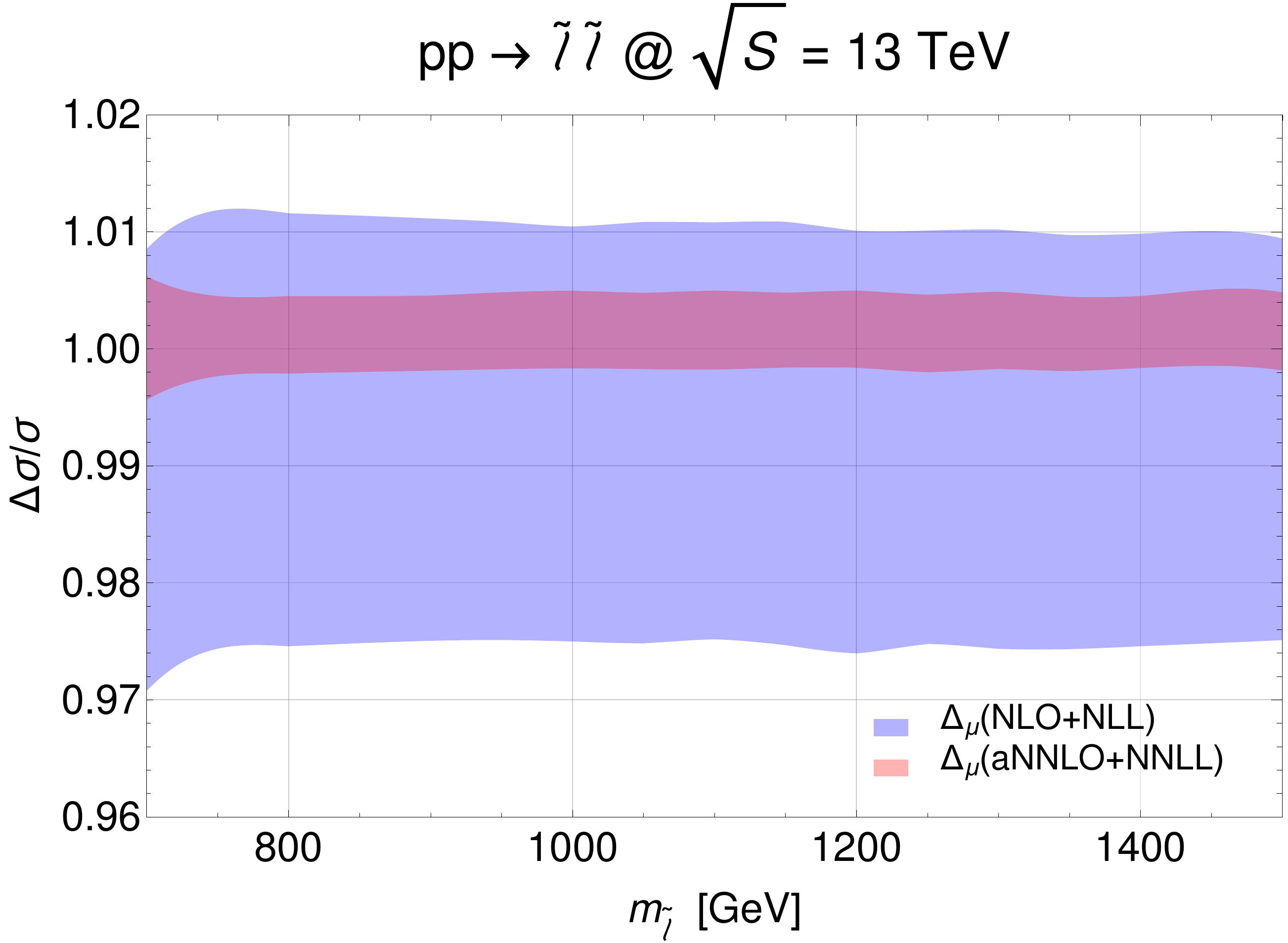}
\caption{(Left) Total cross section for the pair production of left-handed selectron as function of the slepton mass and $K$ factors, and (Right) its scale uncertainty at NLO+NLL and aNNLO+NNLL.}
\label{fig:Sleptons_XS}
\end{figure}

The left plot in Fig.~\ref{fig:Sleptons_XS} shows the slepton pair cross section at different fixed orders and with resummation, as function of the slepton mass, while the panel in the bottom shows the relative K-factors.
The size of the aNNLO+NNLL corrections with respect to the previous NLO+NLL calculation is of the order of 1-2\% in the considered slepton mass range.
In the plot on the right we can appreciate the reduction of the scale uncertainty of the total cross section which shrinks from 1-2\% at NLO+NLL to 0.1-0.2\% at aNNLO+NNLL, uniformly in the whole considered slepton mass range.

\section{Electroweakino pair production at aNNLO+NNLL}

In this section we show the results for the production of neutralinos and charginos in two different realizations of the MSSM.
With the appropriate choice of the model parameters~\cite{Fuks:2017rio}, we generated consistent spectra where the electroweakinos are mostly composed of either Higgsinos or gauginos.

\subsection{Mostly Higgsino electroweakinos}

We consider first the production of electroweakinos in their mostly Higgsino configuration.
In this realization of the MSSM, the lightest supersymmetric particle (LSP), the neutralino $\tilde{\chi}_1^0$, the next-to-lightest neutralino $\tilde{\chi}_2^0$ and the the lightest charginos $\tilde{\chi}_1^\pm$ are almost degenerate in mass, with a mass splitting of the order of 5 - 10 GeV.
The most sensitive exclusion limits for this scenario from ATLAS and CMS are obtained analysing the data set corresponding to an integrated luminosity of 139 fb$^{-1}$, and they set the masses of the considered SUSY particles above roughly 200 GeV~\cite{Aad:2019qnd,CMS:2021xji}.

\begin{figure}
\centering
\includegraphics[width=0.43\textwidth]{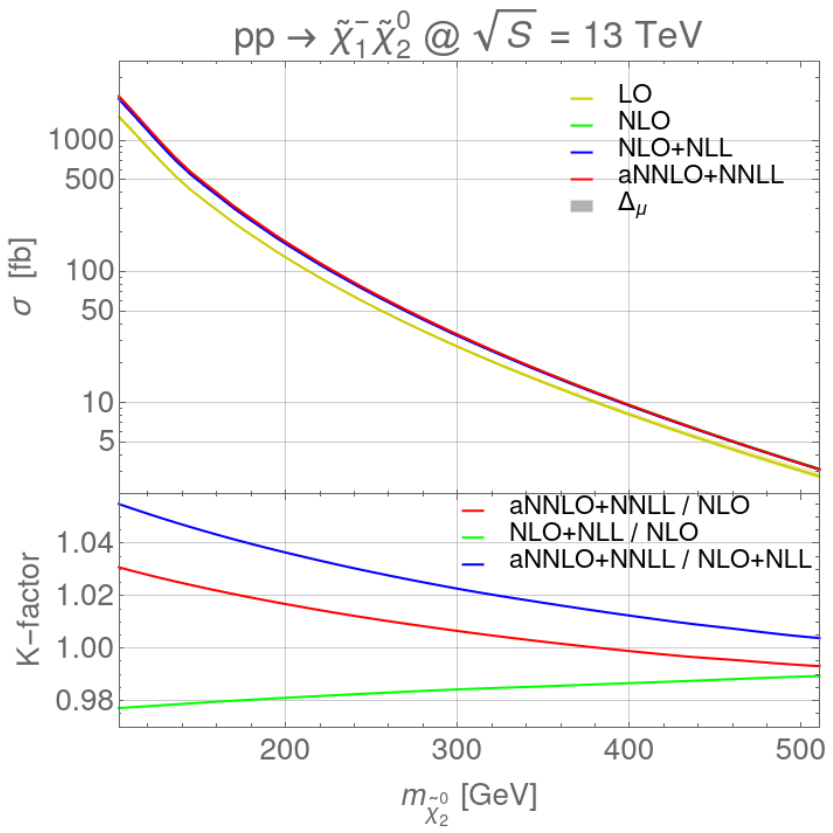}
\includegraphics[width=0.56\textwidth]{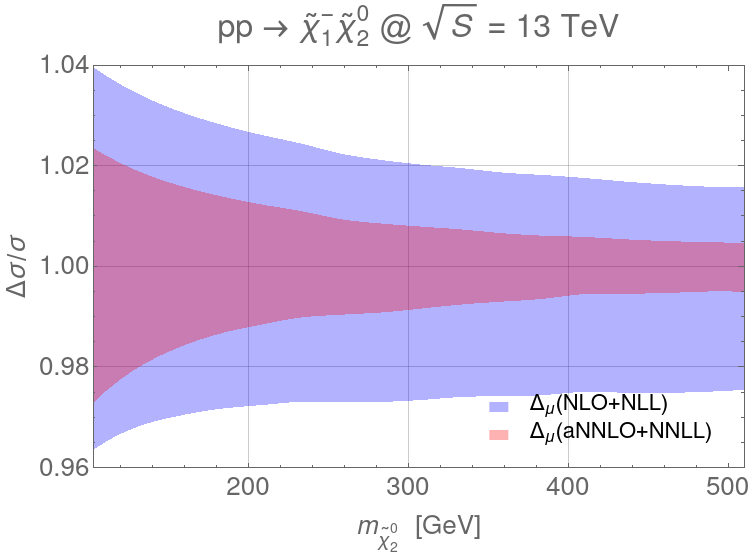}
\caption{(Left) Total cross section for the production of higgsino-like negatively charged chargino and next-to-lightest neutralino as function of the chargino mass and $K$ factors, and (Right) its scale uncertainty at NLO+NLL and aNNLO+NNLL.}
\label{fig:Higgsinos_XS}
\end{figure}

Fig.~\ref{fig:Higgsinos_XS} on the left shows the cross section for the production of a neutralino $\tilde{\chi}_2^0$ and a negatively charged chargino $\tilde{\chi}_1^-$, as function of the neutralino mass, at various fixed orders and with resummation.
The panel in the bottom contains the corresponding K-factors and highlights the sizeable corrections obtained going from NLO+NLL to aNNLO+NNLL calculations, which vary between 1\% and 5\%.
In the plot on the right we observe a significant reduction of the scale uncertainty of the results, which is reduced from about 4\% (2\%) at NLO+NLL down to 1.5\% (0.5\%) for light (heavy) $\tilde{\chi}_2^0$ neutralinos.

\subsection{Mostly gaugino electroweakinos}

Here we consider the scenario of mostly gaugino electroweakinos.
In this configuration the masses of the neutralino $\tilde{\chi}_2^0$ and the charginos $\tilde{\chi}_1^\pm$ are at the TeV scale and almost degenerate, while the LSP $\tilde{\chi}_1^0$ is lighter ($\mathcal{O}$(100) GeV mass).
This choice satisfy the constrains from experimental exclusions by ATLAS and CMS analysis which exclude charginos lighter than 1.0 and 1.4 TeV respectively~\cite{Aad:2019vnb,CMS:2021bra}.

\begin{figure}
\centering
\includegraphics[width=0.43\textwidth]{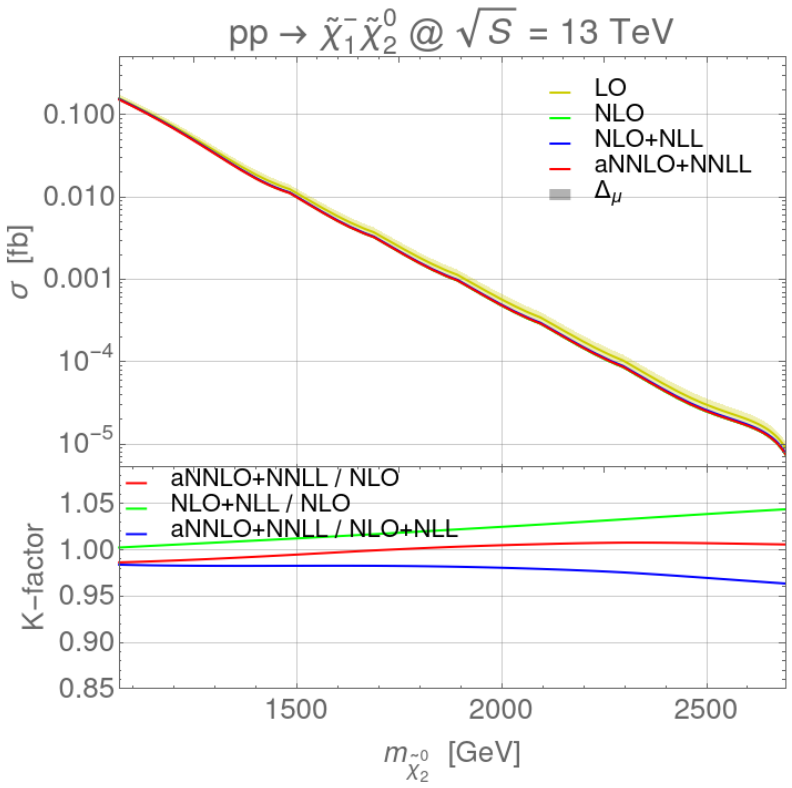}
\includegraphics[width=0.56\textwidth]{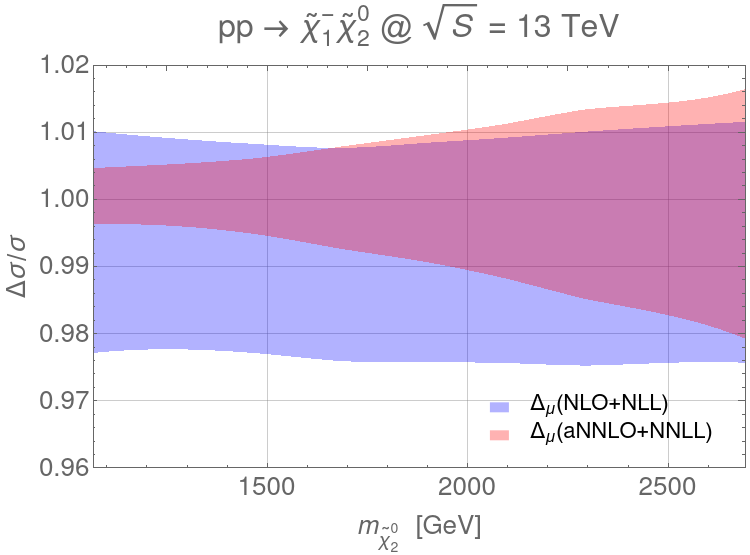}
\caption{(Left) Total cross section for the production of gaugino-like negatively charged chargino and next-to-lightest neutralino as a function of the chargino mass and $K$ factors, and (Right) its scale uncertainty at NLO+NLL and aNNLO+NNLL.}
\label{fig:Gauginos_XS}
\end{figure}

Fig.~\ref{fig:Gauginos_XS} contains the results for the cross section in this scenario in a similar form as above.
Since the particle involved are heavier than the previous case, the size of the radiative corrections is somewhat smaller.
This is visible in the panel containing the K-factors, which shows that the effect of the updated aNNLO+NNLL results on the previous NLO+NLL results is a negative correction of the order of 0.1-0.4\%.
The plot on the right shows a good reduction of the cross section scale uncertainty for light electroweakinos, which goes from 1-2\% at NLO+NLL to 0.5\% at aNNLO+NNLL, while for heavier electroweakinos it remains substantially unchanged.

\section{Conclusions}

The presented results for the electroweak production of sleptons and electroweakinos in their mostly Higgsino and mostly gaugino configurations have been obtained with improved precision at the level of aNNLO+NNLL, for a c.o.m. energy of 13 TeV and with mass scan ranges relevant for current and future LHC analysis.
The impact of the new calculations on the cross section is generally moderate, leading to modification of the order of 1\% or less, however the results feature a sensible reduction of the scale uncertainty which is now around the permil level or below.
The improved calculations have been included in a new release of the RESUMMINO package, now available for public download at \url{https://resummino.hepforge.org/}, which can be employed in future experimental analyses for SUSY searches at the LHC.

\section*{Acknowledgments}
\noindent
This work has been supported by the BMBF under contract 05H18PMCC1 and the DFG through the Research Training Network 2149 ``Strong and weak interactions - from hadrons to dark matter''.
The work of JF has been supported by STFC under the Consolidated Grant ST/T000988/1.

\end{document}